\documentclass[sigconf]{acmart}
\usepackage{tikz}
\usepackage{comment}
\AtBeginDocument{%
  }

\copyrightyear{2026}
\acmYear{2026}
\setcopyright{cc}
\setcctype{by}
\acmConference[CHI EA '26]{Extended Abstracts of the 2026 CHI Conference on Human Factors in Computing Systems}{April 13--17, 2026}{Barcelona, Spain}
\acmBooktitle{Extended Abstracts of the 2026 CHI Conference on Human Factors in Computing Systems (CHI EA '26), April 13--17, 2026, Barcelona, Spain}
\acmDOI{10.1145/3772363.3799032}
\acmISBN{979-8-4007-2281-3/2026/04}

\begin{document}

\title{Alignment--Process--Outcome:\\Rethinking How AIs and Humans Collaborate}

\author{Haichang Li}
\orcid{0009-0006-0952-0709}
\email{hli52@gmu.edu}
\affiliation{%
  \institution{George Mason University}
  \city{Fairfax}
  \state{Virginia}
  \country{USA}
}

\author{Anjun Zhu}
\orcid{0009-0001-9583-8900}
\email{aza99@sfu.ca}
\affiliation{%
  \institution{Simon Fraser University}
  \city{Burnaby}
  \state{British Columbia}
  \country{Canada}
}

\author{Arpit Narechania}
\orcid{0000-0001-6980-3686}
\email{arpit@ust.hk}
\affiliation{%
  \institution{The Hong Kong University of Science and Technology}
  \city{Hong Kong S.A.R.}
  \country{China}
}
\renewcommand{\shortauthors}{Li, Zhu, and Narechania}

\begin{abstract}
In real-world collaboration, alignment, process structure, and outcome quality do not exhibit a simple linear or one-to-one correspondence: similar alignment may accompany either rapid  convergence or extensive multi-branch exploration, and lead to different results. Existing accounts often isolate these dimensions or focus on specific participant types, limiting structural accounts of collaboration.

We reconceptualize collaboration through two complementary lenses. The task lens models collaboration as trajectory evolution in a structured task space, revealing patterns such as advancement, branching, and backtracking. The intent lens examines how individual intents are expressed within shared contexts and enter situated decisions. Together, these lenses clarify the structural relationships among alignment, decision-making, and trajectory structure.

Rather than reducing collaboration to outcome quality or treating alignment as the sole objective, we propose a unified dynamic view of the relationships among alignment, process, and outcome, and use it to re-examine collaboration structure across Human–Human, AI–AI, and Human–AI settings.
\end{abstract}



\begin{teaserfigure}
  \includegraphics[width=\textwidth]{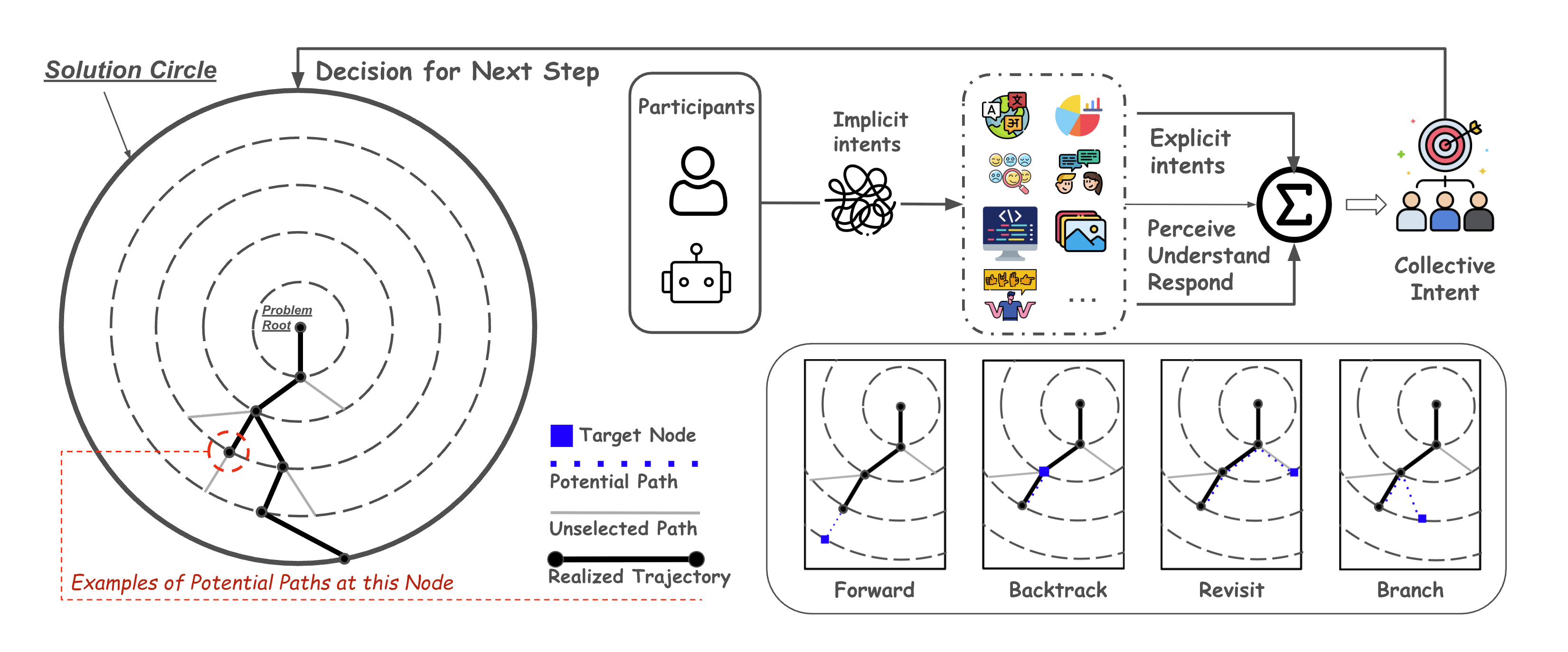}
   \caption{Overview of our two-lens view of collaboration flow. Left: the task lens models collaboration as a trajectory evolving in a task space from the Problem Root toward the Solution Circle. Top-right: the intent lens shows how participants’ implicit intents become explicit and aggregate into a collective intent. Bottom-right: examples of potential trajectory updates at a local decision point in the task space (e.g., forward progression, backtracking, revisiting prior alternatives, and branching).}
   \Description{A three-part conceptual diagram of collaboration flow. Left: a circular task space showing a trajectory from the Problem Root to the Solution Circle. Top-right: an intent lens pipeline where participants’ implicit intents are expressed through interaction channels and aggregated into a collective intent. Bottom-right: multiple possible local trajectory updates, including forward progress, backtracking, revisiting earlier alternatives, and branching into different paths.}
   \label{fig:flow}
\end{teaserfigure}

\maketitle

\section{Introduction}
In real-world settings, alignment, collaboration process, and task outcome often exhibit a persistent disconnect rather than a stable monotonic positive relationship. This means that perfect alignment does not inherently guarantee a smooth, linear process or a superior outcome, as these three dimensions are not naturally coupled. Alignment or misalignment can emerge in collaboration with multi-branch exploration or near-linear progress, and both can lead to strong or weak outcomes. For example, a team that initially struggles to understand each other may still reach a high-quality solution through iterative detours and exploration, while a highly aligned team may progress smoothly yet converge prematurely to a merely average solution.

This disconnect reveals a critical gap: without a unified framework to understand the structural relationships among these three dimensions, research and practice tend to adopt fragmented perspectives, creating problematic blind spots. For example, alignment-centric views risk mistaking consensus for quality, overlooking premature convergence to mediocre solutions. Process-centric views may penalize exploratory backtracking as inefficiency, failing to distinguish productive iteration from genuine waste. Outcome-centric views can obscure whether intent participation was equitable or whether collaboration was merely nominal.

Meanwhile, collaboration research has expanded from traditional Human--Human interaction to AI--AI and Human--AI settings. Yet the underlying theories remain fragmented by participant type and scope of analysis, contributing to these siloed perspectives. Collaboration is often discussed separately across different participant configurations \cite{tran2025multiagentcollaboration,10.1145/3745238.3745531,Yousefi2025team,fragiadakis2025}. Macro-level work often characterizes collaboration through stage models or task categories \cite{tuckman1965,mcgrath1984,doublediamond}, whereas micro-level work focuses on local interaction mechanisms and behavioral details, such as grounding and conversational repair \cite{clark1991grounding,repair1977}. These lines of work are valuable on their own. However, they lack a shared structural discussion to explain how individual intents aggregate and diverge during collaboration, and how they gradually shape the macro-level trajectory of task progress. 

Accordingly, we argue for the need to rethink how humans and AIs collaborate. Some classic theories have also been explicitly framed as descriptive abstractions that generalize empirical patterns across different work domains \cite{Schmidt1996}. Rather than following an empirical path that induces patterns from records of completed collaborations and treats the process as a static object for retrospective analysis, we call for a complementary foundational perspective that derives the evolution of collaboration from the underlying dynamics of the process itself. This perspective starts from task structure and intent aggregation, conceptualizing collaboration as a continuously evolving dynamic process—termed collaboration flow. By examining the dynamics and trajectories of collaboration flow, we re-examine the structural relationships among alignment, collaboration, and task outcomes, providing a more generative starting point for future research and system design.

\section{Two Lenses for Rethinking Collaboration}


We introduce two complementary lenses for observing collaboration: \emph{task lens} and \emph{intent lens}. The \emph{task lens} focuses on the overall evolution from start to finish within the task structure. The \emph{intent lens} moves from individual to collective, examining how intents are expressed, perceived, and aggregated in collaboration. Together, these lenses analyze collaboration flow and support subsequent discussion of the structural relationships among alignment, collaboration, and task outcomes.

\subsection{Task Lens: From Start to Finish}
On the macro level, the task lens treats collaboration as a holistic process, examining its overall evolutionary trajectory from start to finish within the task structure. 

From the task structure perspective, collaboration often begins with a well-defined problem, yet potential solutions and evolutionary paths can take many different forms, making them difficult to characterize by a single dimension. To describe this overall evolution, we abstract the task structure as a \emph{task space} that captures collaboration from start to finish: a 2D-Disk centered at the \emph{Problem Root}, with its boundary defined as the \emph{Solution Circle}.
This disk abstraction stems from a natural deduction of the task's expansive nature: since all collaborative efforts originate from a single root and radiate toward infinite potential outcomes, the task space is inherently radiating in nature. Furthermore, because every final solution represents a 100\% completion state regardless of its specific form, these endpoints are logically equidistant from the root in terms of progress, naturally constituting a circular boundary---the solution circle. 

Collaboration is thus characterized as an evolutionary trajectory from the problem root toward the solution circle. Following a polar-coordinate intuition, the radial direction represents \emph{collaboration depth}, indicating the degree to which the collaboration state advances from the problem root ($d=0\%$) toward the solution circle ($d=100\%$). In this coordinate system, the radial progression directly reflects the reduction in distance to the final goal and solution. Radial progression need not increase monotonically over time; collaboration may stall or involve backtracking, revisiting earlier branches and partially undoing progress. The angular direction corresponds to \emph{path variation}, capturing path choices that lead to differences in final solution forms at similar collaboration depth. This reflects exploration across different branches of the task space.

We illustrate this 2D-Disk presentation in Fig.~\ref{fig:flow} to provide a shared coordinate system, where the radial direction represents \emph{collaboration depth} and the angular direction captures \emph{path variation}. The concentric circles illustrated in the figure serve as conceptual markers provided solely for abstraction and clarity rather than fixed temporal milestones. This distinction accounts for the fact that, in practice, the actual increments of progress resulting from collaborative decisions are often irregular and non-linear, representing discrete movements across different levels of progression within this continuous radial space. We also use this representation to support the interface design of our future trajectory visualization work in Section~4.3.

Under this lens, task completion is determined by radial position, while the final solution is determined by angular path choices; the intersection of the trajectory with the solution circle corresponds to the final solution. Trajectory features (such as backtracking frequency, angular deviation, and branching count) provide observables for discussing collaboration efficiency and exploration.

\subsection{Intent Lens: From Individuals to Collective}
While the trajectory patterns in the task lens (e.g., backtracking frequency or angular shifts) macroscopically reflect collaborative efficiency and exploration, this perspective obscures the composition of local states. Specifically, it masks implicit and conflicting individual goals, as well as potential paths that remain unselected. To address this, we introduce the intent lens as a complementary micro-level analysis to examine how the collaboration flow is constructed from individual to collective intents.

Whether human or AI, participants possess both directly expressed \emph{Explicit Intents} and \emph{Implicit Intents} that cannot be directly perceived. The latter includes unexpressed goals, preferences, and tacit knowledge---internal constraints that continuously influence judgment and decision-making. To function in collaboration, implicit intents must be projected into the shared context through multimodal semantic channels such as language and visualization, becoming explicit intents that others can perceive, understand, and respond to. This process is context-dependent: the representation of intentions in semantic space depends not only on their expressed content but also on the surrounding context. For example, ``I'm okay with that'' may be interpreted as explicit support or reluctant compliance depending on the collaborative situation. Consequently, relative relationships between intentions are often difficult to compare directly across contexts.

Conceptually, intents can be represented in a computational form. These semantic intents (e.g., natural language, gestures, expressions) can be represented via embeddings and compared through similarity metrics  \cite{inproceedings, 10.1145/3680528.3687677}. However, similarity alone is insufficient to characterize intent relationships in collaboration. For instance, ``I support Plan A'' and ``I oppose Plan A'' may be semantically similar yet exert opposite effects on collaboration progress \cite{hasan-ng-2013-frame}. Therefore, representing how individual intents aggregate into collective decisions requires encoding the \emph{relative stance} between intents beyond similarity \cite{hardalov-etal-2022-survey}, reflecting differences in progress direction.

Collaboration can thus be viewed as a series of decision points triggered by intent relationships within local task contexts. At these nodes, intentions are perceived, compared, and weighed, jointly determining whether the collaboration flow advances, branches, or backtracks. The accumulation of local decisions forms the overall trajectory of the flow.

The task lens takes the collective as the unit of analysis, characterizing the overall collaboration trajectory within the task structure. The intent lens focuses on how individual intentions aggregate into collective decisions, thereby driving the flow of collaboration. Neither lens imposes restrictions on entities, together forming an analytical language for understanding the nature of collaboration across Human--Human, AI--AI, and Human--AI scenarios.

\section{Theoretical Deductions and Preliminary Findings}

This section summarizes a set of preliminary understandings regarding the relationships among alignment, the collaboration process, and task outcomes. These findings are derived from a \textbf{natural deduction} of our framework, where the \emph{intent lens} defines the prerequisites for interaction and the \emph{task lens} provides the geometric dimensions of the collaboration space. 

The logic of our derivation is straightforward: since the intent lens requires a ``shared context'' for sensemaking and the task lens defines ``radial'' and ``angular'' directions, alignment naturally emerges as a three-level relation (Contextual, Radial, and Angular). Furthermore, because individual intents aggregate into a ``collective intent'' through a weighting process---where ignoring a viewpoint is functionally equivalent to assigning it a weight of zero---the trajectory's movement becomes a direct manifestation of this weighted decision-making. Finally, because the radial and angular directions in a polar coordinate system are orthogonal and independent, we deduce a structural decoupling between process and outcome. Ultimately, this framework allows us to re-characterize the fundamental interplay between alignment (the modulator), collaboration flow (the execution), and outcome (the destination).

\subsection{Alignment as a Multi-Level Relation}

Recent work suggests that framing alignment as a binary state of “whether aligned” is insufficient \cite{10.1145/3613905.3650948,Terry2023InteractiveAA,INGSTRUP2021267,shen2025positionbidirectionalhumanaialignment}. Through the joint application of the two lenses, alignment emerges as a multi-layered relationship operating at different levels. Building on the introduced radial–angular task lens, and the context-dependence of intent sensemaking, we note three levels of alignment: Contextual, Radial, and Angular, characterizing their distinct roles in collaboration.

\emph{Contextual Alignment} characterizes whether participants' intents reside within the same shared context and can be correctly understood by one another. When contextual alignment is insufficient, misunderstanding blocks effective interaction even if response motivation or stance association exists, preventing collaboration. 

Building on this foundation, \emph{Radial Alignment} governs whether collaboration can continuously advance toward task completion. When participants diverge on the direction of radial progress, collaboration flow may stall or backtrack, reducing or even resetting collaboration depth. Even when intents are semantically highly related, negative stance can offset radial progress and trigger depth regression, reintroducing previously compressed angular path choices. 

When both contextual and radial alignment hold, the focus of alignment shifts to \emph{Angular Alignment}, which characterizes the relative relationships between intents in path selection, shaping the structural form of collaboration flow. When intents are highly related with aligned stance, collaboration tends to advance radially. When intents are related but stance diverges, angular tension is more likely to translate into branching and parallel exploration rather than radial backtracking. For intents with weak or ambiguous stance, highly related intents typically enrich existing paths through information supplementation, while weakly related intents are more likely to introduce new focal points (e.g., shifting from technical implementation to user experience goals).

From this perspective, contextual alignment determines if collaboration can begin, radial alignment affects if it can sustain progress, and angular alignment shapes the path structure in the task space. This layered view explains why similar alignment levels can yield different collaboration trajectory shapes and outcomes.

\subsection{Collaboration as a Weighted Decision-Making Process}
Whether collaboration sustains progress depends on how multiple intents are weighted and aggregated at critical decision points, and alignment primarily modulates the cost and stability of this weighting process. From the intent lens, advancement, backtracking, and branching can be understood as decision outcomes produced by intent weighting. \emph{Weight allocation} mechanisms determine which intents enter decision-making and their degree of influence, shaping the dynamics and path structure of collaboration. Considering common organizational constraints, we preliminarily distinguish between \emph{structured allocation} and \emph{negotiated allocation}, corresponding to different roles of alignment.

In practice, weighting does not occur in a vacuum. In \emph{structured weight allocation}, weight generation is strongly constrained by established agency or protocols (e.g., high-agency individuals, voting, or rotation rules). Collaboration may thus advance despite insufficient alignment, though it risks degenerating into nominal collaboration---unless low-weight intents substantially expand or reshape the collective decision space. In contrast, \emph{negotiated weight allocation} leaves weights unlocked; participants retain decision contribution rights and spontaneously form weighting schemes. Alignment becomes critical for stable weight convergence. Misalignment prevents convergence, forcing teams to invest extra costs in establishing shared context and reaching consensus on depth and path choices. Collaboration flow becomes prone to costly oscillations or deadlock.

Once weights form, multiple intents collapse into a single \emph{collective intent} that updates the trajectory; different weighting regimes can therefore yield different exploration patterns and land on different regions of the solution circle.

\subsection{Rethinking the Relationship Between Alignment, Collaboration Process, and Outcome}

As previously discussed, collaboration process and task outcome are structurally decoupled: outcome depends on which path variations are retained or compressed during collaboration, not on whether the trajectory is tortuous or linear. The key lies in how weight allocation aggregates multiple intents into collective intent and determines branch retention at critical nodes, causing collaboration to land in different regions of the solution circle. Section 3.1 further discusses how misalignment exerts compounded effects on this mechanism by altering weight convergence and path retention.

Building on this, we examine cases where all three alignment levels hold: collaboration often exhibits rapid and nearly linear radial progression. However, high efficiency does not necessarily guarantee high-quality outcomes. Since angular divergence may be systematically compressed, collaboration flow can still prematurely converge to local optima or structurally biased regions. Conversely, moderate intent divergence, while increasing coordination costs, may introduce path tension and trigger weight reallocation, enabling the collaboration trajectory to cover a broader solution space.

Based on this preliminary understanding, we view alignment as a variable modulating collaboration cost and stability, weight allocation as the execution mechanism driving collaboration forward, and outcome as the landing point of the collaboration trajectory in the solution space. Completing a collaboration merely means the trajectory reaches the solution circle, without guaranteeing optimality of its endpoint in task terms.

\section{Proposed Calls and Implications}
Building on the above reasoning, we propose the following calls and describe our work-in-progress system.

\subsection{Revisiting Metrics: Trajectory, Efficiency, and Outcome}

Collaboration quality cannot be adequately proxied by any single signal. Focusing solely on outcome systematically flattens process structure, such as exploration scope, convergence patterns, and whether intents enter critical decisions. Focusing solely on alignment risks misinterpreting low-friction progression as high-quality collaboration, overlooking outcome quality and path coverage. Optimizing only efficiency or exploration may ignore intent relationships and weight aggregation mechanisms, yielding favorable process metrics but endpoints that deviate from goals. We thus call for joint metrics characterizing collaboration by simultaneously considering outcome quality, progression efficiency, and trajectory structural features (advancement, branching, backtracking). Meanwhile, if systems treat backtracking as failure, weight allocation will systematically penalize exploratory intents. Instead, backtracking should be viewed as a low-cost, recordable, reusable collaboration product. Quantifying how intents enter decisions requires collaboration-oriented intent representation: encoding stance direction relative to shared goals beyond semantic similarity, distinguishing semantically similar intents with opposite directional effects, and linking their impact on weight aggregation and trajectory evolution.

\subsection{Rethinking System Strategies: Structural Plasticity and Strategic Friction}
Weights should not be rigidly bound to collaborator identity. We call for treating weight allocation as a context-switchable protocol layer, reversibly transitioning between negotiated and structured allocation: strengthening structural constraints during oscillations to restore progression, and preserving controlled divergence and parallel exploration when convergence risks arise, maintaining path diversity. Under the coupling of alignment and weight allocation, premature convergence is a common structural risk. Early-stage collaboration should thus moderately preserve divergence rather than uniformly suppress it. In certain contexts, introducing moderate friction (e.g., reflection roles or delayed critical confirmations) may prove more effective \cite{10.1145/3449287,10.1007/978-3-031-61353-1_1,10.5555/3604650.3604651}. This further points to an AI design strategy: AI agents should not blindly defer to humans but introduce interpretable strategic friction at critical nodes to mitigate bias and premature convergence risks. As ongoing work, we are exploring agent support strategies for different misalignment scenarios and integrating them with the trajectory visualization interface discussed later.

\subsection{Redesigning Interface Paradigms: From Linear Streams to Topological Navigation}
Collaboration can be understood as trajectory evolution in task space; linear collaboration logs obscure branching, backtracking, and unchosen paths, impeding review, comparison, and diagnosis. We call for shifting collaboration interfaces from linear flows to navigable trajectory and topological representations, enabling users to explicitly examine progression depth, critical branches and backtracking, and understand how intents aggregate into actions at decision nodes \cite{10.1145/3746059.3747746,10.1145/3698061.3726935}. As ongoing work, we explore visualization and interaction mechanisms for collaboration trajectories and examine their embedding in real workflows. For example, in conversational AI IDEs such as Cursor \cite{anysphere_cursor2026}, task phases, critical decision points, and historical exploration paths could be presented as navigable structures, supporting comparison, backtracking, and reuse. At the same time, this interface design is generalizable to a wider range of application contexts, including writing and design.

\section{Discussion and Future Work}
Although the above analysis avoids binding collaboration mechanisms to any specific participant type, we build on these preliminary results to discuss how they manifest differently across collaboration settings and to clarify the structural factors that require additional attention when rethinking how humans and AIs collaborate.

\subsection{Discussions Under Different Participant Type Settings}
In Human--Human collaboration, the weight allocation process is often pre-shaped by hierarchy and social norms~\cite{socialnorm2023,wadhwa2025designing}, such that apparent progress may stem from preset weights rather than meaningful intent participation. For instance, passive deference to a superior person can compress path exploration and induce nominal collaboration and premature convergence. DAOs~\cite{dao2020} can be seen as attempts to decouple weights from identity; however, without stable constraints, negotiated weight allocation can easily turn into high-friction oscillation. Together with large individual capability gaps and substantial intent projection loss~\cite{polanyi1966tacit}, designs that reduce the cost of contextual alignment (e.g., slides~\cite{visualinfo02} or explicit terminology explanations~\cite{axaridou2018vista}) often become essential for stabilizing the collaboration flow.


In AI--AI collaboration, performance differences across agents are typically less diverse. Since social hierarchies do not exist in this case, there are no social anchors to limit the weight allocation process, making weights more likely to degenerate into high-friction negotiation. Thus collaboration flow may constantly oscillate at decision points instead of reaching a stable conclusion, which explains why multi-agent systems are not necessarily better than single-model execution; recent evidence suggests that as base models become more capable, the marginal gains from multi-agent systems can shrink and may even reverse~\cite{kim2025sciencescalingagentsystems}. Systems therefore often introduce structured weights (e.g., manager agents~\cite{wu2024autogen}, workflow orchestration~\cite{openai_agentbuilder_2025}, or protocols~\cite{anthropic_agent_skills_2025}) to restore progress, and rely on cross-task memory~\cite{memory}, explicit reasoning~\cite{openai_o1_2024}, and context engineering~\cite{anthropic_context_engineering_2025} to reduce intent projection loss. Yet overly strong structure can also systematically compress the path space, allowing faster convergence to the solution circle while increasing the risk of landing in biased regions.


In Human--AI collaboration, the weight boundaries are usually set by humans, but in practice they can also be dynamically adjustable: local decisions can be delegated to AI through explicit authorization (e.g., Full Permission in coding agent settings~\cite{anysphere_cursor2026,anthropic_claude_code_2025}); while humans can actively reclaim control when the stakes escalate or when novel, high-risk, or nuanced situations arise~\cite{hu2025designing}. Vibe Coding~\cite{sarkar2025vibecodingprogrammingconversation,karpathy2025tweet} suggests that even under imperfect alignment between human intent and AI execution, structured weights can still sustain progress and yield usable outcomes. Meanwhile, the intent loss during the translation phase from human brain to human-side Human-AI interface is inevitable. For instance, previous work finds that people often fail to construct and iteratively improve effective prompts for coding~\cite{zamfirescu2023johnny}, and the resulted cognitive bias can lead to sub-optimal outcomes~\cite{zhou2026cognitive}. This unavoidable intention loss means that certain methods including carefully designed prompt engineering~\cite{sahoo2025prompt}, intent-augmentation~\cite{promptiverse,intent2025}, and low-barrier AI interaction~\cite{a11yshape}, are necessary to help shape the collaboration trajectory. However, these interventions often work by making weight allocation and intent expression more controllable and inspectable. The central challenge, therefore, is to support interpretable weight switching~\cite{94debate,agency2025,horvitz1999mixed-initiative}--a long-standing HCI debate on user control, agent autonomy, and mixed-initiative interaction. Yet, preventing interfaces from obscuring the collaboration trajectory and AI's intermediate work at the same time else it is difficult to inspect how and why collaboration converged to a particular solution region.

\subsection{Future Work and Downstream Effects}

 Our discussion is primarily conceptual and qualitative. The geometric formulation serves as a structural metaphor rather than a formal characterization. Operationalizing this framework requires further work along multiple directions. To address this limitation, future efforts should transform traditional linear collaboration logs into a topological navigation interface that explicitly visualizes collaboration flow. Such an interface would enable users to inspect and reflect on the collaboration process. To support quantitative analysis, it is necessary to develop intent vector encoding methods that go beyond simple semantic similarity. This includes constructing a computational approach that jointly represents intent content, similarity, and stance relations. We refer to this representation as intent embedding.

 In terms of downstream effects, this framework offers a novel perspective for the HCI community to re-examine how we observe, diagnose, and design collaboration. At the observation level, it shifts attention from isolated, surface-level phenomena to insights into structural patterns. For example, when a team reaches consensus quickly, the framework invites us to use angular coverage of the trajectory to assess whether this reflects efficient progress or premature convergence driven by insufficient exploration of alternatives. At the diagnosis level, the framework turns ambiguous explanations into traceable mechanisms. By characterizing the morphology of collaborative trajectories and conducting statistical analyses at key decision points, we can intuitively describe collaborative performance. For instance, by examining how branching and revisit behaviors relate to alignment at different levels, we can offer actionable recommendations and precisely articulate the health of collaboration. At the design level, the framework points to a shift from single-objective optimization to structured balancing. Designers can preserve angular divergence early to avoid premature convergence. They can introduce strategic friction at critical junctures to sustain path diversity. They can also use switchable weighting protocols to ensure that radial progress does not stagnate. More broadly, this perspective can ground a foundational framework for more comprehensive reasoning about collaboration. It encourages the community to re-examine the structure of collaboration, rather than explaining collaboration solely through outcomes or treating alignment—especially in human–AI settings—as the only objective. This structural approach provides researchers with a more principled empirical basis, system designers with theoretically grounded design guidelines, and practitioners with higher-dimensional awareness and control over collaborative processes.

\section{Conclusion}
We take the position that the relationship among alignment, collaboration process, and task outcome is not a stable, one-dimensional positive correlation. We introduce two analytic lenses—\emph{task} and \emph{intent}—to view collaboration as trajectory evolution within a task, driven by how multiple intents are weighted and aggregated into decisions. Under this view, alignment modulates coordination cost, while weight allocation determines whether exploration is preserved or compressed, shaping both trajectory structure and outcome quality. We hope this perspective can inform future metrics, system strategies, and interface paradigms for collaboration between and among humans and AIs.

\bibliographystyle{ACM-Reference-Format}
\bibliography{references}

\end{document}